\documentclass[runningheads]{llncs}

\usepackage{epsfig}

\def\href#1#2{#2}

\newcommand{\progfont}[0]{\rm\small}
\newcommand{\Upipe}[0]{\(|\)}
\newcommand{\Uoutputannotation}[0]{\(<\!\!<\!\!<\)}
\newcommand{\Urule}[0]{\(\leftarrow\)}
\newcommand{\UIrule}[0]{\(%
  \stackrel{\scriptscriptstyle\infty}{\leftarrow}\)}
\newcommand{\UInrule}{\(
  \not\stackrel{\scriptscriptstyle\infty}{\leftarrow}\)}
\newcommand{\ULongrightarrow}[0]{\(\Longrightarrow\)}
\newcommand{\ULeftrightarrow}[0]{\(\Leftrightarrow\)}
\newcommand{\Unotunify}[0]{\(\backslash\!\!\!=\)}
\newcommand{\Unrule}{\(\not\leftarrow\)}
\newcommand{\Udoublequote}{"{}}

\def\K#1{{%
   \progfont{%
   #1}}} 

\newlength{\davor}
\def\durch#1{%
   \settowidth{\davor}{#1}%
   {\makebox[0pt][l]{#1}%
      \makebox[0pt][l]{\raisebox{.4ex}{\makebox[\davor]{\hrulefill}}}%
      \makebox[0pt][l]{\raisebox{.8ex}{\makebox[\davor]{\hrulefill}}}%
      \raisebox{1.2ex}{\makebox[\davor]{\hrulefill}}%
   }}

\def\unterstrichen#1{%
   \settowidth{\davor}{#1}%
   {\makebox[0pt][l]{#1}%
      \raisebox{-.5ex}{\makebox[\davor]{\hrulefill}}%
   }}

\newcommand{\prologcodefont}[0]{\progfont}
\newcommand{\internneuezeile}[0]{\\[0pt]\prologcodefont}
\newcommand{\interntabulator}[0]{\>\prologcodefont}
\newcommand{\prologcode}[0]{\prologcodefont}

\newenvironment{Pro}[0]
{\quad\begin{minipage}[t]{40ex}%
   \prologcode%
   \begin{tabbing}%
\hspace{3ex}\=\hspace{3ex}\=\hspace{3ex}\=\hspace{3ex}\=%
\hspace{6ex}\=\hspace{6ex}\=\hspace{6ex}\=\hspace{6ex}\=%
\hspace{6ex}\=\hspace{6ex}\=\hspace{6ex}\=\hspace{3ex}\kill}%
{\end{tabbing}%
 \end{minipage}%
}

\begin{document}

\setcounter{page}{77}
\title{Declarative program development in Prolog with GUPU}
\titlerunning{Declarative program development in Prolog with GUPU}
\author{Ulrich Neumerkel \and Stefan Kral}
\authorrunning{U. Neumerkel, S. Kral}
\institute{Institut f{\"u}r Computersprachen\\
           Technische Universit{\"a}t Wien, Austria,\\
           \email{ulrich@complang.tuwien.ac.at skral@complang.tuwien.ac.at}}

\maketitle

\addtocounter{footnote}{1}
\footnotetext{In Alexandre Tessier (Ed), proceedings of the 12th International Workshop on Logic Programming Environments (WLPE 2002), July 2002, Copenhagen, Denmark.\\Proceedings of WLPE 2002: \texttt{http://xxx.lanl.gov/html/cs/0207052} (CoRR)}

\begin{abstract}

We present GUPU, a side-effect free environment specialized for
programming courses.  It seamlessly guides and supports students
during all phases of program development, covering specification,
implementation, and program debugging.  GUPU features several
innovations in this area.  The specification phase is supported by
reference implementations augmented with diagnostic facilities.
During implementation, immediate feedback from test cases and from
visualization tools helps the programmer's program understanding.  A
set of slicing techniques narrows down programming errors.  The whole
process is guided by a marking system.

\end{abstract}

\section*{Introduction}

Teaching logic programming has specific opportunities different from
traditional languages.  In particular, the declarative notions can be
liberated from theoretical confines and be applied to actual program
development.  Our approach is centered around the programming course
environment GUPU used for introductory Prolog programming courses
since 1992 at TU~Wien and other universities.  The language used by
GUPU is the monotone pure subset of Prolog which also contains the
many constraint extensions offered by SICStus Prolog.

In this article, we focus on the program development process supported
by GUPU.  To illustrate this process, we will develop the predicate
\K{alldifferent/1} describing a list of pairwise different elements.

The actual programming effort is divided into two stages which both
are equipped with appropriate diagnosing facilities.  The first stage
(Sect.~1) is devoted to specifying a predicate by example.  Cases are
stated where the predicate should succeed, fail, terminate, or not
terminate.  In the second stage (Sect.~2), the actual predicate is
implemented and immediately tested against the previously stated
example cases.  The whole development process is guided by a marking
system that highlights missing items and computes an interval
percentage of the fulfillment of an exercise.

\section{Specification by example}

Writing test cases prior to actual coding is good practice also in
other programming languages.  It has reached broader attention with
the rise of the extreme programming movement~\cite{XP}.  The major
advantage generally appreciated is increased development speed due to
the following reasons.  Test cases are an unambiguous (albeit
incomplete) specification.  They influence system design, making it
better testable.  They provide immediate feedback during coding, which
is most important in our context.  And finally, they constitute a
solid starting point for documentation.

In the context of logic programming, there are several further aspects
in favor of this approach.  Test cases focus the attention on the
meaning of the predicate avoiding procedural details.  In particular,
recursive predicates are a constant source of misunderstandings for
students, because they confuse the notion of termination condition in
imperative languages with Prolog's more complex mechanisms.  While
there is only a single notion of termination in procedural languages,
Prolog has two: Existential and universal termination.  By writing
tests prior to coding, the student's attention remains focused on the
use of a predicate.  With logic variables, test cases are more
expressive than in imperative languages.  We can use these variables
existentially (in positive queries) as well as universally (in
negative queries).

In our attempt to develop \K{alldifferent/1} describing a list of
pairwise different elements, we start by writing the following two
assertions. The first is a positive assertion.  It ensures that there
must be at least a single solution for \K{alldifferent/1}.  The second
is a negative one which insists that the given goal is not the case.
Here, \K{alldifferent/1} should not be true for a list with its two
elements being equal.  Both tests cannot be expressed in traditional
languages.

\begin{Pro}%
\Urule{} alldifferent(Xs).\internneuezeile
\Unrule{} alldifferent([X,X]).\internneuezeile
{\bf @! Definition of alldifferent/1 missing in above assertions}%
\end{Pro}

\noindent
We write test cases directly into the program text.  Upon saving, GUPU
inserts feedback {\em into} that text.  Above, GUPU added the line
starting with {\bf @} to remind us that we have not yet defined
\K{alldifferent/1}.

Taking all of the aforementioned advantages into account, this
methodology appears preferable to the traditional ``code then test''
approach.  Still, students prefer to write code prior to testing.  It
seems that the biggest obstacle is a lack of motivation.  Test cases
as such do not provide any immediate feedback when they are written.
Why should one write tests when they might be incorrect?  Such
incorrect tests would be misleading in the later development.  For
this reason GUPU tests assertions for validity.

\subsection{Reference implementation}

All test cases provided by the student are tested against a reference
implementation which is realized with otherwise inaccessible
predicates.  It is considered to be correct for those cases where the
reference predicate fails finitely or succeeds unconditionally.  In
all other cases (exceptions, non-termination, pending constraints),
the reference implementation is incomplete and therefore unspecified
as discussed in the sequel.

Cases detected as incompatible with the reference implementation are
highlighted immediately.  In this manner, it is possible to ``test the
tests'' without any predicate yet defined by the student.  Errors due
to the reference implementation start with \K{!=} to distinguish them
from other errors.

We continue with the specification of \K{alldifferent/1}, by adding
further positive and negative assertions.  In the cases below GUPU
disagrees and offers help.

\begin{Pro}%
\Urule{} alldifferent([a,b,c,d,c]).\internneuezeile
{\bf != Should be negative.  Details with DO on the arrow}\internneuezeile
\internneuezeile
\Unrule{} alldifferent([X,Y\Upipe{}\_]).\internneuezeile
{\bf != Should be positive.  Details with DO on the arrow}%
\end{Pro}

\subsection{Diagnosis of incorrect assertions}

A more elaborate explanation on why an assertion is incorrect is
obtained on demand as proposed by the message ``DO on the arrow''
above.  According to the kind of assertion the following steps are
taken.  In case of an incorrect negative assertion, a more specific
query is produced if possible.  For incorrect positive assertions, a
generalized query is given.  In this manner, GUPU provides a detailed
diagnosis without showing the actual code of the reference
implementation.  Moreover, further assertions are offered that will
improve test coverage during coding.

\paragraph{Explaining incorrect negative assertions.}

If the reference implementation succeeds in a negative assertion, a
more specific goal is obtained using an answer substitution of the
reference implementation.  All free variables are grounded with new
constants \K{any1, any2, ...}.

Below, an answer substitution completed the end of the list with
\K{[]}, and the variables \K{X} and \K{Y} have been grounded to
constants.  GUPU temporarily inserts the specialized assertion into
the program text.  By removing the leading @-signs, the assertion can
be added easily to the program.

\begin{Pro}%
\Unrule{} alldifferent([X,Y\Upipe{}\_]).\internneuezeile
{\bf @@ \% != Should be a positive assertion.}\internneuezeile
{\bf @@ \% Also this more specific query should be true.}\internneuezeile
{\bf @@ \Urule{} X = any0, Y = any1, alldifferent([X,Y]).}%
\end{Pro}

\paragraph{Explaining incorrect positive assertions.}
If the reference implementation fails for a positive assertion, GUPU
tries to determine a generalized goal that fails as well.  Generalized
goals are obtained by rewriting the goal (or a conjunction of goals)
up to a fixpoint with the following rules.

\begin{description}

\item[R1:] Replace a goal (in a conjunction) by true.

\item[R2:] Replace a subterm of a goal's argument by a fresh variable
  \K{\_}.

\item[R3:] Replace two or more identical subterms by a new shared
  variable.

\item[R4:] Replace two non-unifiable subterms by new variables V1, V2
  and add the goal \K{dif(V1, V2)}.

\item[R5:] Replace a goal by a set of other goals that are known to
  be implied.

\end{description}

\begin{Pro}%
\Urule{} alldifferent([a,b,c,d,c]).\internneuezeile
{\bf @@ \% != Should be a negative assertion}\internneuezeile
{\bf @@ \% @ Generalized negative assertion: \hfill--- using R1,R2}\internneuezeile
{\bf @@ \Unrule{} alldifferent([\_,\_,c,\_,c]).}\internneuezeile
{\bf @@ \% @ Further generalization: \hfill--- using R1-R4}\internneuezeile
{\bf @@ \Unrule{} alldifferent([\_,\_,V0,\_,V0]).}\internneuezeile
{\bf @@ \% @ Further generalization: \hfill --- using R1-R5}\internneuezeile
{\bf @@ \Unrule{} alldifferent([V0,\_,V0\Upipe{}\_]).}%
\end{Pro}

\noindent
In our experience it is helpful to use several stages of
generalizations.  Explanations that are easier to compute are
presented first.  In the first stage, R1 and R2 are used, and the
generalization is displayed immediately.  In the next stage, R1-R4 may
incur a significant amount of computation.  In particular, R4 is often
expensive if the number of non-unifiable subterms is large.  Note that
the obtained generalized goals are not necessarily optimal because of
the incompleteness of the reference implementation.  In the above
example, the first generalization is sub-optimal.  The optimum for R1
and R2 is \K{\Unrule{} alldifferent([\_,\_,c,\_,c\Upipe{}\_]).} However, our
reference implementation loops in this case.  In the last
generalization, R5 applied ``\K{alldifferent([\_\Upipe{}Xs]) \ULongrightarrow{}
  alldifferent(Xs).}'' twice.  This permitted R1 to remove constant
\K{[]} at the end of the list.

\subsection{Incomplete reference implementations}

For most simple predicates, our reference implementation is capable of
determining the truth of all simple assertions.  There are, however,
several situations where no feedback is provided.

\begin{itemize}
\item The reference implementation takes too long, although most
  reference predicates take care of many situations and are therefore
  more complex than the student's code.

\item The predicate itself is under-specified.  Many under-specified
  predicates, however, still allow for partial assertion testing.

\end{itemize}

\noindent
The former situation has been illustrated previously with
\K{alldifferent([\_,\_,c,\_,c\Upipe{}\_])}.  For the latter, consider a
family database, the usual introductory example.  While the particular
persons occurring in \K{child\_of/2} and \K{ancestor\_of/2} are not
fixed, there still remain many constraints imposed on them.

\begin{enumerate}
  \item[c1] No child has three parents.
  \item[c2] A parent is an ancestor of its children.
  \item[c3] The ancestor relation is irreflexive.
  \item[c4] The ancestor relation is transitive.
\end{enumerate}

We give here the actual reference implementation of \K{child\_of/2}
and \K{ancestor\_of/2} implemented in CHR~\cite{CHR}, a high-level
language to write constraint systems with simplification (\ULeftrightarrow{}) and
propagation rules (\ULongrightarrow{}).  The definition provides two predicates
that can be executed together with regular predicates of the reference
implementation.  This reference implementation cannot succeed
unconditionally because of the pending constraints imposed by CHR.
Therefore, it can only falsify positive assertions.

\begin{Pro}%
\% {\bf Reference implementation in CHR}\internneuezeile
\Urule{} use\_module(library(chr)).\internneuezeile
option(already\_in\_store, on). \% prevents infinite loops\internneuezeile
\internneuezeile
c1 @ child\_of(C,P1), child\_of(C,P2), child\_of(C,P3) \ULeftrightarrow{}\internneuezeile
{\interntabulator}{\interntabulator}{\interntabulator}true \& P1 \Unotunify{} P2, P2 \Unotunify{} P3, P1 \Unotunify{} P3 \Upipe{} false.\internneuezeile
c2 @ child\_of(A,B) \ULongrightarrow{} ancestor\_of(B,A).\internneuezeile
c3 @ ancestor\_of(A,A) \ULeftrightarrow{} false.\internneuezeile
c4 @ ancestor\_of(A,B), ancestor\_of(B,C) \ULongrightarrow{} ancestor\_of(A,C).\internneuezeile
\internneuezeile
\Urule{} child\_of(A,B), child\_of(B,C), A = C.\internneuezeile
{\bf != Should be negative.}\internneuezeile
\Urule{} alldifferent([P1,P2,P3]), child\_of(C,P1), child\_of(C,P2), child\_of(C,P3).\internneuezeile
{\bf != Should be negative.}%
\end{Pro}

\subsection{Termination}

Termination properties and in particular non-termination properties
are often considered unrelated to the {\em declarative} meaning of a
program.  It is perceived that ideally a program should always
terminate.  We note that non-termination is often closely related to
completeness.  In fact, a query {\em must not} terminate if the
intended meaning can only be expressed with an infinite number of
answer substitutions.  Notice that this observation is completely
independent of Prolog's actual execution mechanism!  No matter how
sophisticated an execution mechanism may be, its termination property
is constrained by the size of the generated answer.  If this answer
must be infinite, non-termination is inevitable.  For this reason,
cases of non-termination that are due to necessarily infinite sets of
answer substitutions can be safely stated in advance.  On the other
hand, cases of termination always depend on the particular predicate
definition as well as on Prolog's execution mechanism.

The most interesting cases of termination are those where actual
solutions are found.  Termination is therefore expressed with two
assertions: A positive assertion to ensure a solution: \K{\Urule{} Goal.}
A negative one to ensure universal termination under Prolog's
simplistic left-to-right selection rule: \K{\Unrule{} Goal, false.}  In
our particular case of \K{alldifferent/1}, we can state that
\K{alldifferent(Xs)} must not terminate, because there are infinitely
many lists as solutions.  With the negative infinite assertion
\K{\UInrule{} Goal, false.} we state that \K{Goal} must not terminate
universally.

\begin{Pro}%
\Urule{} alldifferent(Xs).\internneuezeile
\UInrule{} alldifferent(Xs), false.%
\end{Pro}

When the length of the list is bounded (at most) a single answer
substitution for \K{alldifferent/1} is possible.  Therefore, the
predicate could terminate if defined appropriately.  In the example
below, the ideal answer is \K{dif(A,B)}.

\begin{Pro}%
\Urule{} Xs = [A,B], alldifferent(Xs).\internneuezeile
\Unrule{} Xs = [A,B], alldifferent(Xs), false.%
\end{Pro}

\subsection{Summary}
In the first stage of realizing a predicate, only test cases are
given.  The idea is to start from the most general goal and refine the
meaning of the predicate by adding further cases.  To streamline this
process, their correctness is ensured with the help of an internal
reference implementation.  It provides immediate feedback and detailed
diagnosis in the form of further test cases that can be added to the
program.  Without any actual predicate code written, the learner is
put into the situation of formulating and reading queries generated by
the system.  Note that in this first stage most errors remain local in
each query.  The marking system provides overall guidance by demanding
various forms of assertions.  E.g., a ground positive assertion,
non-termination annotations, etc.

\section{Predicate definition}

Armed with validated test cases obtained in the first stage, we are
now ready to implement \K{alldifferent/1}.  Inconsistencies between
the student's test cases and implementation, are highlighted
immediately.  Answer substitutions are tested with the reference
implementation, as shown in the central column of the following table.
Explanations based on slicing locate an error.  We continue our
example by defining the predicate with an error in the underlined
part.

\begin{Pro}%
alldifferent([]).\internneuezeile
alldifferent([X\Upipe{}Xs]) \Urule{}\internneuezeile
{\interntabulator}nonmember\_of(\unterstrichen{\bf Xs, X}),\internneuezeile
{\interntabulator}alldifferent(Xs).%
\end{Pro}
\begin{Pro}%
nonmember\_of(\_X, []).\internneuezeile
nonmember\_of(X, [E\Upipe{}Es]) \Urule{}\internneuezeile
{\interntabulator}dif(X, E),\internneuezeile
{\interntabulator}nonmember\_of(X, Es).%
\end{Pro}

\begin{Pro}%
\Urule{} X = any1, Y = any2, alldifferent([X,Y]).\internneuezeile
{\bf ! Unexpected failure.  Explanation with DO on the arrow.}\internneuezeile
\Urule{} Xs = [\_,\_], alldifferent(Xs).\internneuezeile
{\bf != The\,first solution is incorrect.  Explanation with DO on\,the arrow.}\internneuezeile
\Unrule{} Xs = [\_,\_], alldifferent(Xs), false.\internneuezeile
{\bf ! Universal non-termination.  Explanation with DO on the arrow.}%
\end{Pro}

\vspace{-3ex}

\subsection{Slicing}

Slicing~\cite{slice-use} is a technique to facilitate the
understanding of a program.  It is therefore of particular interest
for program debugging.  The basic idea of slicing is to narrow down
the relevant part of a program text.  For the declarative properties
insufficiency (unexpected failure) and incorrectness (unexpected
success), two different slicers have been realized.  A further slicer
was realized to explain universal non-termination~\cite{uwn-ppdp}.
They all highlight fragments of the program where an error has to
reside.  As long as the programmer does not modify the displayed
fragment, the error persists.

\begin{enumerate}

\item[a)] For unexpected failures, generalized program fragments are
  produced that still fail.  The program is generalized by deleting
  some goals, indicated with a *-sign.  To remove the error, the slice
  must be generalized.

  \quad E.g., a rule \K{p \Urule{} q, r.} is generalized to \K{p \Urule{} *
    \durch{q}, r.}

\item[b)] In case of unexpected success, still succeeding specialized
  fragments are obtained by inserting some goals \K{false/0} that
  effectively remove program clauses.  In order to remove the error,
  the programmer has to specialize the remaining program fragment.

\item[c)] For universal non-termination, slicing determines still
  non-terminating specialized fragments.  The inserted
  \K{false/0}-goals hide all subsequent goals in a clause.  If
  \K{false/0} is inserted as the first goal, the clause is completely
  eliminated.  The remaining program fragment has to be modified in
  order to remove non-termination.  The constraint based algorithm is
  found in~\cite{uwn-ppdp}.

  \quad E.g., \K{p \Urule{} p, q.} is specialized to \K{p \Urule{} p,
  false, \durch{q}.}

\end{enumerate}

\noindent
GUPU generates the following slices on demand (``DO on the
arrow'').

\noindent
\renewcommand{\progfont}[0]{\rm\scriptsize}
\begin{tabular}{l|l|l}
\hspace{-4ex}
\begin{Pro}%
\Urule{} X= any1, Y= any2,\internneuezeile
{\interntabulator}{\interntabulator}{\interntabulator}alldifferent([X,Y]).\internneuezeile
{\bf ! Unexpected failure.}\internneuezeile
\internneuezeile
\internneuezeile
\internneuezeile
\internneuezeile
{\bf ------ Explanation, ad a) ---}\internneuezeile
Generalized fragment fails.\internneuezeile
\internneuezeile
\Urule{} X= any1, Y= any2,\internneuezeile
{\interntabulator}{\interntabulator}{\interntabulator}alldifferent([X,Y]).\internneuezeile
alldifferent([]).\internneuezeile
alldifferent([X\Upipe{}Xs]) \Urule{}\internneuezeile
{\interntabulator}nonmember\_of(\progfont\unterstrichen{\progfont\bf Xs, X}),\internneuezeile
{\interntabulator}* \durch{alldifferent(Xs)}.\internneuezeile
\internneuezeile
nonmember\_of(\_X, []).\internneuezeile
nonmember\_of(X,[E\Upipe{}Es]) \Urule{}\internneuezeile
{\interntabulator}* \durch{dif(X, E)},\internneuezeile
{\interntabulator}* \durch{nonmember\_of(X, Es)}.%
\end{Pro}
&\hspace{-3ex}
\begin{Pro}%
\Urule{} Xs = [\_,\_], alldifferent(Xs).\internneuezeile
{\bf @@ \%  Xs = [[],[]].\,\%\,Incorrect!}\internneuezeile
{\bf @@ \% Generalization}\internneuezeile
{\bf @@ \Unrule{} alldifferent([[],[]\Upipe{}\_]).}\internneuezeile
{\bf @@ \%\,Further\,generalization}\internneuezeile
{\bf @@ \Unrule{} alldifferent([V0,V0\Upipe{}\_]).}\internneuezeile
\internneuezeile
{\bf ------ Explanation, ad b) ------{}--}\internneuezeile
Specialized fragment succeeds.\internneuezeile
\internneuezeile
\Unrule{} alldifferent([V0,V0\Upipe{}\_]).\internneuezeile
\internneuezeile
alldifferent([]).\internneuezeile
alldifferent([X\Upipe{}Xs]) \Urule{}\internneuezeile
{\interntabulator}nonmember\_of(\unterstrichen{\bf Xs, X}),\internneuezeile
{\interntabulator}alldifferent(Xs).\internneuezeile
\internneuezeile
nonmember\_of(\_X, []).\internneuezeile
\durch{nonmember\_of(X, [E\Upipe{}Es]) \Urule{} false,}\internneuezeile
{\interntabulator}\durch{dif(X, E),}\internneuezeile
{\interntabulator}\durch{nonmember\_of(X, Es).}%
\end{Pro}
&\hspace{-3ex}
\begin{Pro}%
\Unrule{} Xs = [\_,\_], alldifferent(Xs),\internneuezeile
{\interntabulator}{\interntabulator}{\interntabulator}{\interntabulator}{\interntabulator}{\interntabulator}false.\internneuezeile
{\bf !\,Universal\,non-termination }\internneuezeile
\internneuezeile
\internneuezeile
\internneuezeile
\internneuezeile
{\bf ------ Explanation, ad c) --{}--}\internneuezeile
Fragment does not terminate.\internneuezeile
\internneuezeile
\Unrule{} Xs = [\_,\_], alldifferent(Xs),\internneuezeile
{\interntabulator}{\interntabulator}{\interntabulator}{\interntabulator}{\interntabulator}{\interntabulator}false.\internneuezeile
\durch{alldifferent([]) \Urule{} false.}\internneuezeile
alldifferent([X\Upipe{}Xs]) \Urule{}\internneuezeile
{\interntabulator}nonmember\_of(\unterstrichen{\bf Xs, X}), false,\internneuezeile
{\interntabulator}\durch{alldifferent(Xs)}.\internneuezeile
\internneuezeile
\durch{nonmember\_of(\_X, []) \Urule{} false.}\internneuezeile
nonmember\_of(X, [E\Upipe{}Es]) \Urule{}\internneuezeile
{\interntabulator}dif(X, E),\internneuezeile
{\interntabulator}nonmember\_of(X, Es), false.%
\end{Pro}
\end{tabular}
\renewcommand{\progfont}[0]{\rm\small}

\noindent
In the above example, two declarative errors and a procedural error
yielded three different explanations.  All explanations exposed some
part of the program where an error must reside.  Reasoning about
errors can be further enhanced by combining the obtained explanations.
Under the assumption that the error can be removed with a single
modification of the program text, the error has to reside in the
intersection of the three fragments.  The intersection of all three
fragments comprises only two lines of a total of eight lines --- the
head and first goal of \K{alldifferent/1}.  This intersection is
currently not generated by GUPU.

\begin{Pro}%
\durch{alldifferent([]).}\internneuezeile
alldifferent([X\Upipe{}Xs]) \Urule{}\internneuezeile
{\interntabulator}nonmember\_of(\unterstrichen{\bf Xs, X}),\internneuezeile
{\interntabulator}\durch{alldifferent(Xs)}.%
\end{Pro}
\begin{Pro}%
\durch{nonmember\_of(\_X, []).}\internneuezeile
\durch{nonmember\_of(X, [E\Upipe{}Es]) \Urule{}}\internneuezeile
{\interntabulator}\durch{dif(X, E),}\internneuezeile
{\interntabulator}\durch{nonmember\_of(X, Es).}%
\end{Pro}

The advantages of using slicing in the context of a teaching
environment are manifold.  The students' attention remains focused on
the logic programs and not on auxiliary formalisms like traces.
Reading program fragments proves to be a fruitful path toward program
understanding.  Simpler and smaller parts can be read and understood
instead of the complete program.  Further, the described techniques
are equally used for monotone extensions of pure Prolog programs like
constraints in the domain CLP(FD).

\subsection{Beyond Prolog semantics}

The truth of infinite queries cannot be tested with the help of
Prolog's simplistic but often efficient execution mechanism.  GUPU
subjects infinite queries implicitly to an improved execution
mechanism based on iterative deepening.  In contrast to approaches
that exclusively rely on iterative deepening~\cite{yal}, we can
therefore obtain the best of both worlds: Prolog's efficiency and a
sometimes more complete search.  In the case of negative infinite
queries, a simple loop checking prover is used.  We are currently
investigating to further integrate more sophisticated techniques.

\begin{Pro}%
nat(s(N)) \Urule{}\internneuezeile
{\interntabulator}nat(N).\internneuezeile
nat(0).\internneuezeile
\UInrule{} nat(N).\internneuezeile
{\bf !+$\!\!$+ Unexpected\,success.\,Remove\,/}
\end{Pro}
\ 
\begin{Pro}%
q \Urule{}\internneuezeile
{\interntabulator}q.\internneuezeile
\internneuezeile
\UIrule{} q.\internneuezeile
{\bf !+$\!\!$+ Unexpected\,failure.\,Add\,/}%
\end{Pro}

\subsection{Viewers}

In a pure logic programming environment, the only way to see the
result of a computation is via answer substitutions which often serves
as an excuse to introduce impure features and side-effects.  In
GUPU, answer substitutions are visualized in a side-effect free manner
with the help of viewers.  Viewers provide an alternate visual
representation of an answer substitution which helps to understand the
investigated problem.  To ensure that no side effects take place,
viewers are only allowed in assertions of the form \K{\Urule{} {\em
    Viewer} \Uoutputannotation{} {\em Goal}.} When querying such an assertion, an
answer substitution of {\em Goal} is displayed along with a separate
window for the viewer. {\em Viewer} is one of the predefined viewers.
Most complex viewers are based on the Postscript viewer which expects
a string describing a Postscript document.  In fact, many viewers have
been implemented side-effect free within GUPU in student projects.  We
present two select viewers.  Further viewers are discussed
in~\cite{uwn-viewers}.

\begin{Pro}%
\Urule{} postscript(Cs) \Uoutputannotation{} Cs = \Udoublequote{}0 0 100 100 rectfill\Udoublequote{}.\internneuezeile
{\bf @@ \%  Cs = \Udoublequote{}0 0 100 100 rectfill\Udoublequote{}.}\internneuezeile
{\bf @@ \%\% One solution found.}%
\end{Pro}

\paragraph{Repetitive Scheduling.} Within the context of fault-tolerant,
distributed hard real-time systems, the need to calculate time-rigid
schedules arises.  Before system start up, a schedule meeting the time
criteria of the application is calculated.  The resulting table is
executed at run time by the components of the system.  This real-world
application of CLP(FD)~\cite{AE} has been integrated into GUPU.  The
viewer \K{repsched/1} displays one particular solution to this
scheduling problem: \\\K{\Urule{} repsched(DB-S) \Uoutputannotation{} DB = big,
  db\_timerigidschedule(DB,S).}

The modeled system consists of several processors and of one global
interconnect (a bus). During design-time, every task is assigned to
exactly one processor.  All tasks are executed periodically and are
fully preemptive.  Tasks communicate and synchronize by sending
messages.  Internal messages are used for tasks on the same processor.
All other messages are sent over the bus.

\noindent
\begin{minipage}[b]{0.49\linewidth}
Solutions to the scheduling problem obey several constraints. All
messages must be transmitted after the completion of the sending task
and before starting the receiving task.  The bus transfers at most one
message at a time.  Transactions (specific groups of tasks) have a
guaranteed maximal response time (time from the start of the earliest
task to the end of the last completing task).  Tasks having a period
smaller than the period of their transaction are replicated
accordingly.  Processors execute at most one task at a time.

\end{minipage}
\hfill
\begin{minipage}[t]{.45\linewidth}
\epsfig{file=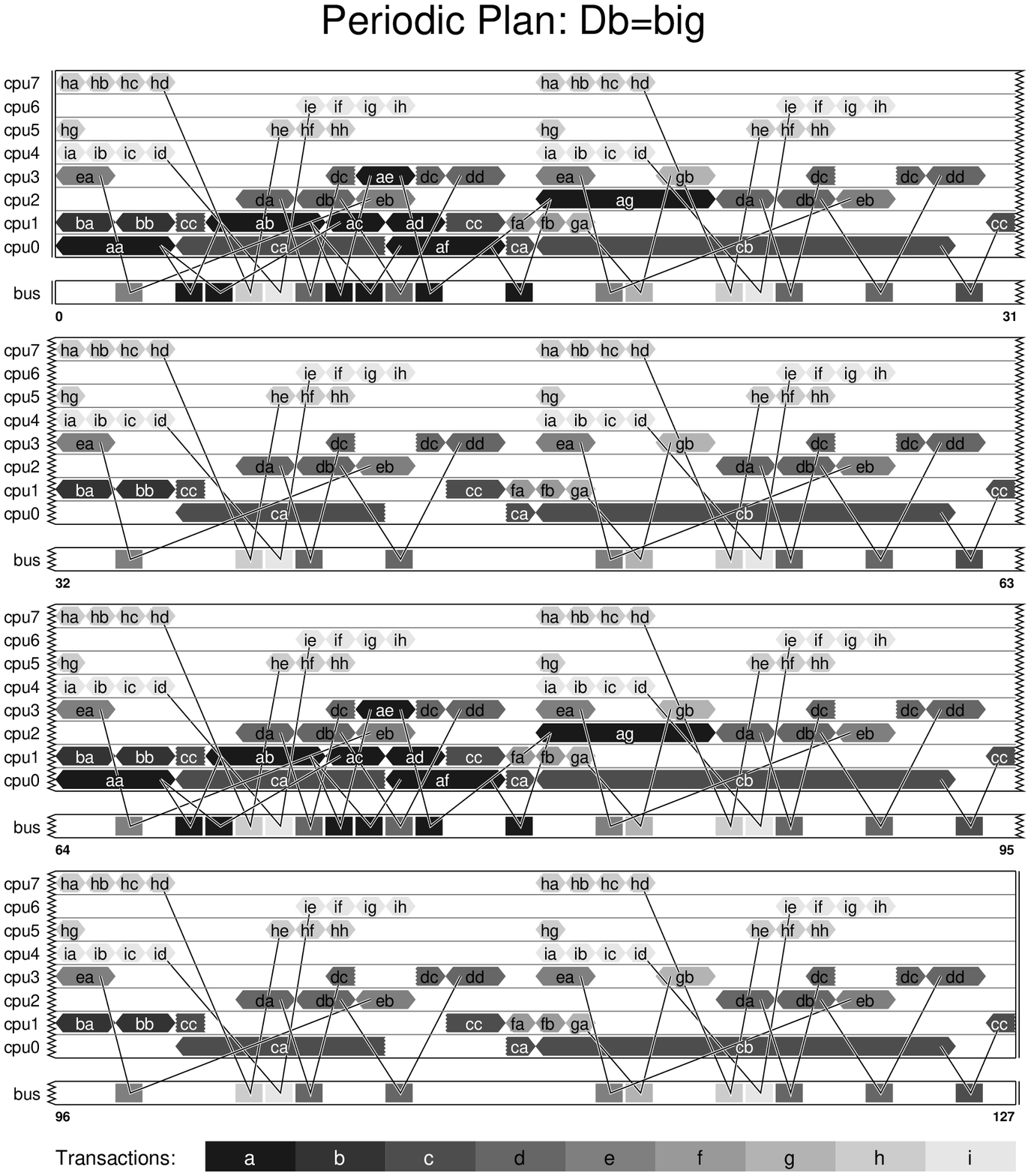,width=\linewidth}
\end{minipage}

\paragraph{An agent environment.}

Another critical issue apart from basic I/O concerns the side-effect
free representation of logical agents.  As an example, the wumpus cave
has been realized, well known from introductory AI texts~\cite{aima}.
In this world the agent is supposed to find and rescue gold in a dark
cave guarded by a beast and paved with other obstacles.  Only by using
rudimentary perception, the agent makes its way through the cave.  The
basic functionality of the agent is represented with two predicates.
One for initialization of the agent's state \K{init(State)} and one to
describe the agent's reaction upon a perception
\K{percept\_action\_(Percept, Action, State0, State)}.  In addition,
the state of the agent's knowledge can be communicated via
\K{maybehere\_obj\_(Position, Object, State)} which should succeed if
the agent believes at the current \K{State} that an \K{Object} may be
located at \K{Position}.

\begin{center}
\begin{minipage}[t]{.48\linewidth}
\epsfig{file=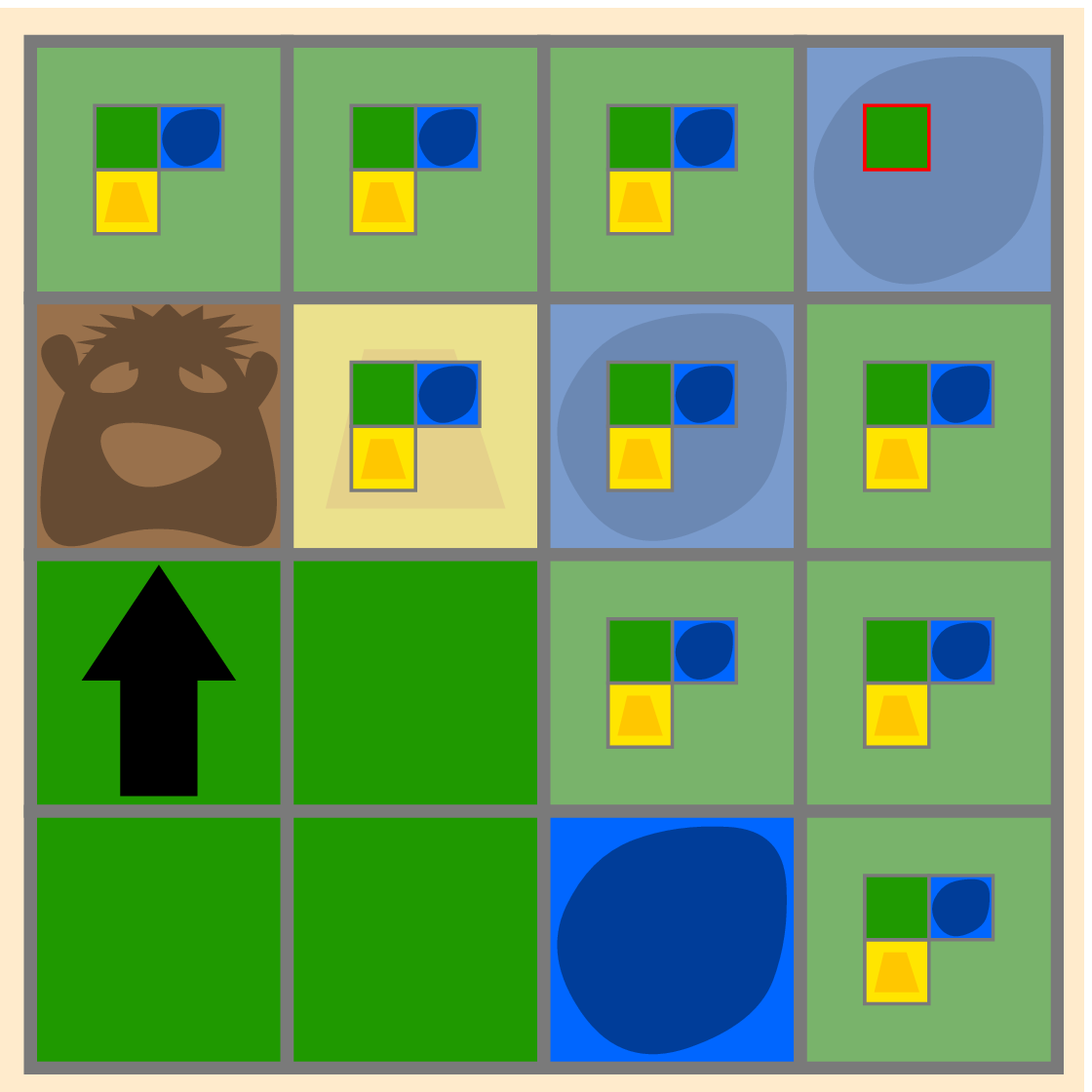,width=\linewidth}
\end{minipage}
\hfill
\begin{minipage}[t]{.42\linewidth}
\epsfig{file=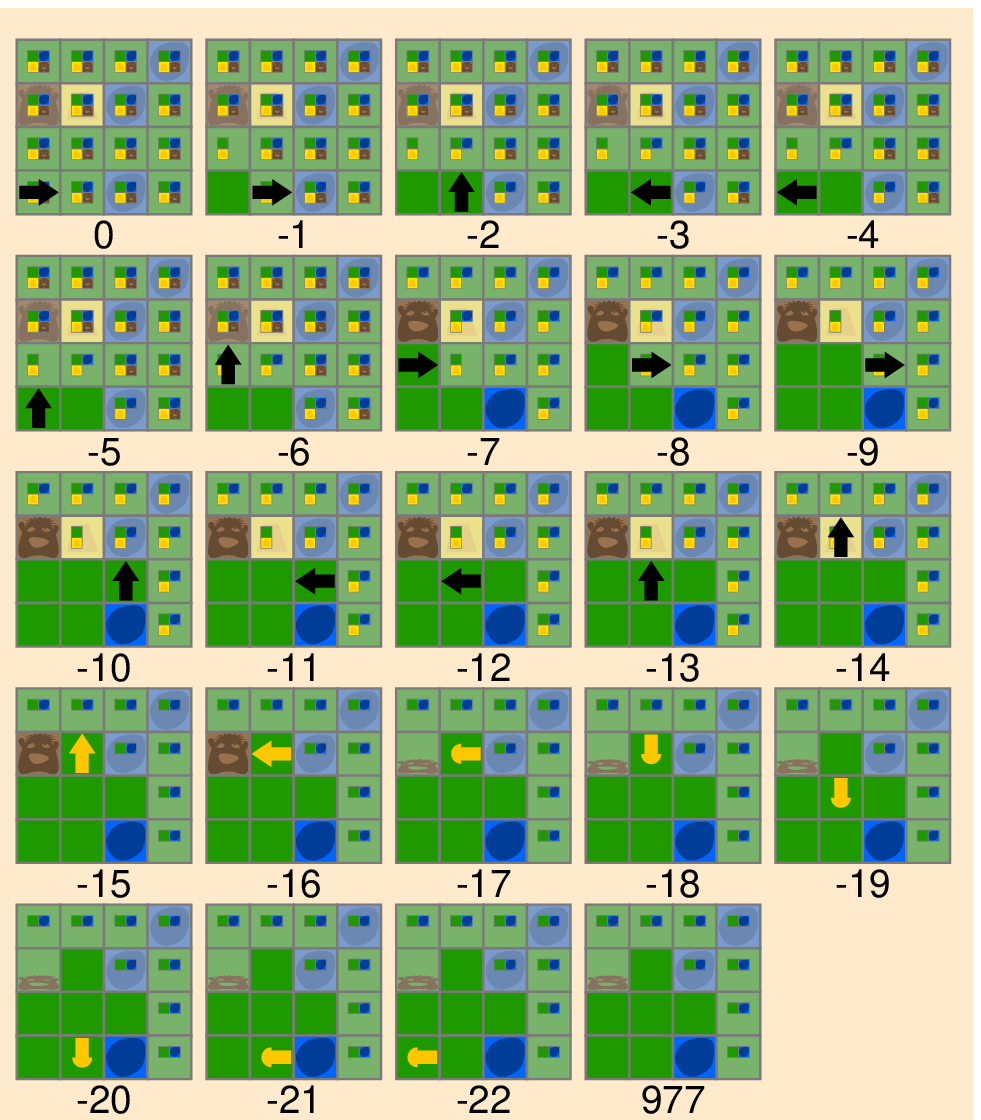,width=\linewidth}
\end{minipage}
\end{center}

The graphical representation shows the agent depicted as an arrow
walking through the cave as well as the location of the objects
invisible to the agent.  The agent's belief is represented by the
centered squares on each field.  The agent may believe that a field is
free, a pit, contains gold, or is occupied by the wumpus.  In the
picture, the agent has discovered the wumpus, a pit, and some free
fields.  All other fields are believed to contain a free field, gold,
or a pit.  Inconsistencies between the agent's belief and reality are
highlighted as depicted by the field in the upper right corner.  While
that field is a dangerous pit, the agent believes it to be safe.  The
left viewer only displays a single situation at a time, the other
viewer presents the complete course through the cave at a glance,
simplifying the comparison of different agents.

\paragraph{Related Work.}
Ushell~\cite{yal} uses iterative deepening in introductory courses.
GUPU resorts to better strategies only when Prolog takes too long.
CIAOPP~\cite{ciaopp} provides a rich assertion language (types, modes,
determinacy, cost, ...).  It is much more expressive than GUPU's but
also complex to learn.  Prolog~IV~\cite{prologIV} has a very sophisticated
assertion language which is particularily well suited for constraints.
Approaches to teaching Prolog with programming
techniques~\cite{techniques} highlight patterns otherwise invisible to
the inexperienced.  Advice is given on the programming technique and
coding level~\cite{hong}.  GUPU provides guidance prior to any coding
effort.  We believe programming techniques might also be useful for
GUPU.

\paragraph{Acknowledgments.}
Our thanks go to all our tutors, Heimo Adelsberger, and the Austrian
Computer Society (OCG).  This work was partially supported by the
German Bundesministerium für Bildung und Forschung.  Anton Ertl, David
Gregg, and Franz Puntigam provided valuable comments.

\end{document}